\documentclass[aps,pra,reprint,superscriptaddress,showpacs]{revtex4-1}
\usepackage{graphicx}
\usepackage{amsmath}
\usepackage{textcomp}
\usepackage{color}
\usepackage{xcolor}
\usepackage{upgreek}
\usepackage{verbatim}
\usepackage{amssymb}
\usepackage{tabu}
\usepackage{float}
\usepackage[english]{babel}
\usepackage{gensymb}

\begin{document}

\preprint{yes}

\setlength{\abovedisplayskip}{5pt}
\setlength{\belowdisplayskip}{5pt}

\title{Mimicking Chiral Light-Matter Interaction}


\author{Sergey Nechayev}
\email[]{sergey.nechayev@mpl.mpg.de}
\affiliation{Max Planck Institute for the Science of Light, Staudtstr. 2, D-91058 Erlangen, Germany}
\affiliation{Institute of Optics, Information and Photonics, University Erlangen-Nuremberg, Staudtstr. 7/B2, D-91058 Erlangen, Germany}





\author{Peter Banzer}
\affiliation{Max Planck Institute for the Science of Light, Staudtstr. 2, D-91058 Erlangen, Germany}
\affiliation{Institute of Optics, Information and Photonics, University Erlangen-Nuremberg, Staudtstr. 7/B2, D-91058 Erlangen, Germany}

\date{\today}

\begin{abstract}
We demonstrate that electric-dipole scatterers can mimic chiral light-matter interaction by generating far-field circular polarization upon scattering, even though the optical chirality of the incident field as well as that of the scattered light is zero. The presented effect originates from the fact that electric-dipole scatterers respond selectively only to the incident electric field, which eventually results in depolarization of the transmitted beam and in generation of far-field circular polarization. To experimentally demonstrate this effect we utilize a cylindrical vector beam with spiral polarization and a spherical gold nanoparticle positioned on the optical axis -- the axis of rotational symmetry of the system. Our experiment and a simple theoretical model address the fundamentals of duality symmetry in optics and chiral light-matter interactions, accentuating their richness and ubiquity yet in highly symmetric configurations.
\end{abstract}
\maketitle
\section{Introduction} 
Chiral light-matter interactions attract tremendous attention in modern classical~\cite{Kopp74,bliokh_geometrodynamics_2008,chiral_backsc_2011,valev_chirality_2013,cameron_robert_p._chirality_2017} and quantum optics~\cite{Zhang725, zoller2015, arno_isolator_2015,Lodahl2017}. Describing chiral light-matter interactions requires characterization of chirality of the incoming and outgoing electromagnetic field. Optical chirality~\cite{Tang2010} is closely related to helicity~\cite{Bliokh2011_02, Bliokh2013, Nieto2015}, a quantity that measures whether the total angular momentum is aligned or anti-aligned with linear momentum~\cite{Ivan2012,Cameron_helicity2012, Ivan2013}. The helicity density $K \propto \Im \left( \mathbf{E}^{\ast} \cdot \mathbf{H}\right)$ in the far-field is proportional to the degree of circular polarization of each individual plane-wave, expressed via the angularly resolved third Stokes parameter $S_3(\mathbf{k})$, where $\mathbf{k}$ is the wavevector. In focused fields, $K$ is proportional to a difference between the integrated contributions of all right- and left-hand circularly polarized (RCP and LCP) plane-waves to the focal field~\cite{Ivan2012,Bliokh2013,Cameron_helicity2012, Ivan2013, Bliokh2011_02, Bliokh2014_07,Nieto2015, Nieto2017, Nieto2017_05}. Owing to the fundamental relation between light's chirality and helicity density $K$, extinction of helicity from an incident field in the course of light-matter interaction is usually a clear signature of interaction of matter and light chirality~\cite{Bliokh2014_07,Bliokh2014_07,Ivan_chirality, Nieto2015, Nieto2017, Nieto2017_05}. Selective extinction or generation of helicity by chiral molecules upon interaction with the incident light was first observed by Haidinger in 1847, and it was termed \textit{circular dichroism} by Cotton in 1895 and became the most important tool in identifying chiral objects since then~\cite{barron_2004}.\\
Here, we present a counter-intuitive case of interaction between an electric-dipole scatterer and achiral light in cylindrically symmetric configuration that results in generation of circular polarization, effectively mimicking an interaction of matter and light chirality. Specifically, we show that scattering of a focused cylindrical vector beam with spiral polarization~\cite{spiral_beam,Eismann2018} by an electric-dipole scatterer generates far-field helicity. Notwithstanding that the incident beam, the focal fields, as well as the scattered light bear zero optical chirality and the scatterer is positioned on the optical axis -- the axis of rotational symmetry of the system. Moreover, the whole experimental system is cylindrically symmetric and independent of the direction of incidence of the beam~\cite{Zambrana2014,zambrana-puyalto_far-field_2016,Revah:18}. The physical origin of the effect is related to the fact that electric-dipole scatterers break the electromagnetic duality symmetry~\cite{Ivan2012,Cameron_helicity2012,Ivan2013,Bliokh2013,Nieto2015} by responding selectively only to the incident electric field components. Based on this, the experiment can be explained in terms of the superposition of the transverse electric (TE or azimuthal) component of the incident beam and the phase-delayed transverse magnetic (TM or radial) component of the scattered light. Our experiment and a simple theoretical model shed light on chiral light-matter interactions and helicity conservation laws in electromagnetism.
\section{Theory} 
\begin{figure}[htb!]
  \includegraphics[width=0.5\textwidth]{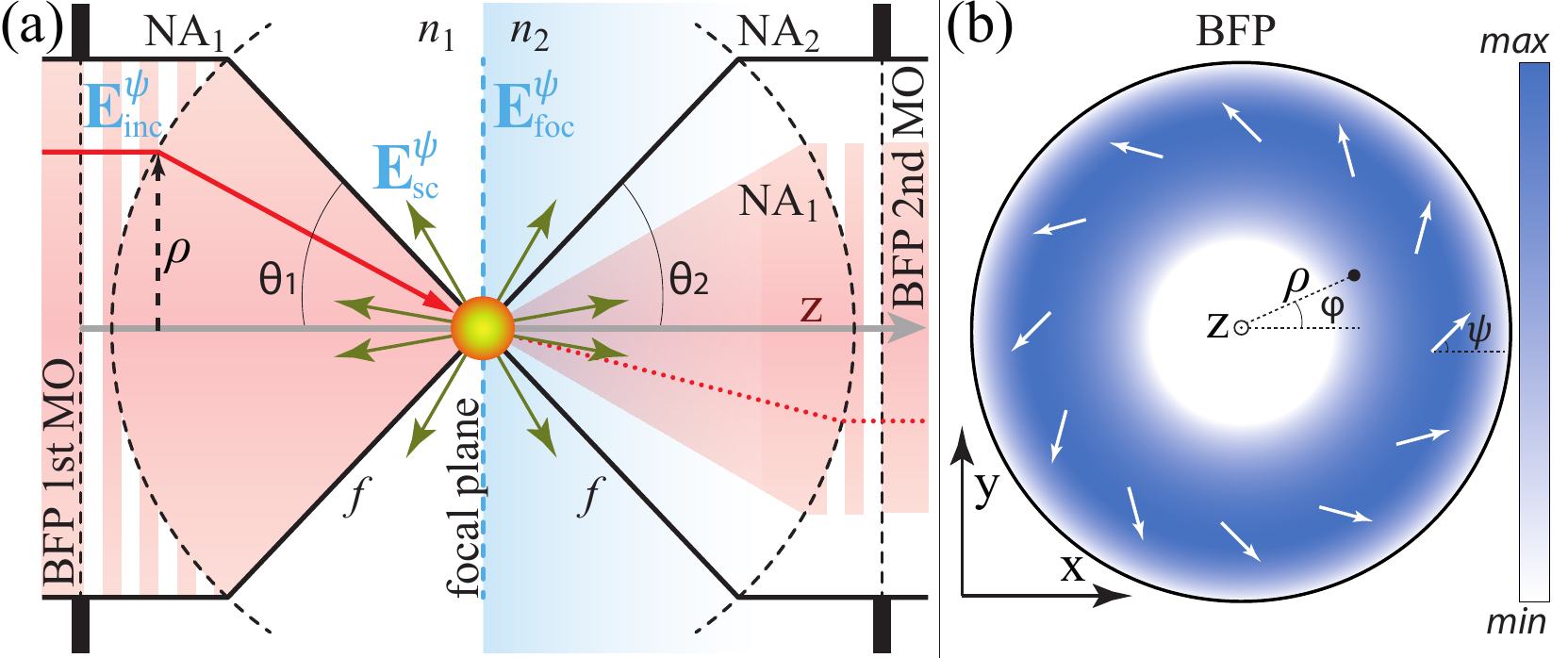}
  \caption{(a) An aplanatic high numerical aperture (NA) microscope objective (MO) focuses the incident field distribution in its back focal plane (BFP) $\mathbf{E}_{\text{inc}}^\psi$. The focal plane ($z=0$) constitutes a boundary between two dielectric media with refractive indices $n_1$ and $n_2$. The focal field $\mathbf{E}_{\text{foc}}^\psi$ excites a spherical electric-dipole-like nanoparticle positioned along the optical axis $z$ at the position $z=z_0$. The second confocally aligned index-matched aplanatic MO collimates the incident beam and collects the scattered light $\mathbf{E}_{\text{sc}}^\psi$. The interference pattern of $\mathbf{E}_{\text{inc}}^\psi$ and $\mathbf{E}_{\text{sc}}^\psi$ is observed in the BFP of the second MO. The theoretical description is given for the case of $n_1=n_2=1$, NA$_1$ = NA$_2 = 1$ and $z_0=0$. In the experimental section we use $n_1=1$, $n_2=1.52$, NA$_1=n_1 \max\left\{  \sin( \theta_1 ) \right\}=0.9$, NA$_2=n_2 \max\left\{\sin(\theta_2 ) \right\}=1.3$ and $z_0=-d/2=70\,\mathrm{nm}$, where $d=140\,\mathrm{nm}$ is the diameter of the nanoparticle. (b) Distribution of the incident spirally-polarized cylindrical vector beam in the BFP of the first MO. The intensity pattern is shown as the colormap, while the white arrows depict the polarization pattern.}
  \label{fig:focusing}
\end{figure}

For the theoretical treatment, we consider a system consisting of two confocally aligned aplanatic microscope objectives (MO) with focal lengths $f$ and numerical apertures $\mathrm{NA}=\mathrm{NA}_1=\mathrm{NA}_2$ with the surrounding refractive index $n=n_1=n_2=1$, as shown in Fig.~\ref{fig:focusing}(a). The incident spirally-polarized cylindrical vector beam (SPCVB)~\cite{spiral_beam}, schematically shown in Fig.~\ref{fig:focusing}(b), propagating along the $z$-axis is focused by the first MO, while the second MO collects and collimates the transmitted light. The incident field distributions in the back focal planes (BFP) of both MOs are given by:
\begin{align} \begin{split}
&\mathbf{E}_{\text{inc}}^\psi \equiv E_{\text{in}}\mathbf{e}_{\text{in}}=E_{\text{in}}    ( \cos(\psi) \hat{\boldsymbol{\rho}} +\sin(\psi) \hat{\boldsymbol{\varphi}})\\
&\mathbf{H}_{\text{inc}}^\psi \equiv \frac{ E_{\text{in}} }{\eta}\mathbf{h}_{\text{in}}=\frac{ E_{\text{in}} }{\eta} (\cos(\psi) \hat{\boldsymbol{\varphi}} - \sin(\psi) \hat{\boldsymbol{\rho}})
\text{,}\\
\end{split} \label{eq:inc} \end{align}
with a doughnut-shaped amplitude profile \mbox{$E_{\text{in}}=E_0\frac{\rho}{w_0}\exp\left(- \frac{\rho^2}{w_0^2}\right) $}, where $w_0$ is the beam waist, $\eta$ is the freespace impedance, $\rho$ and $\varphi$ are the radial and axial cylindrical coordinates, respectively, $\psi$ defines the spiral polarization angle, and $E_0 = 1$ without the loss of generality. Additionally, aplanatic MOs link the field distributions in their BFPs to the far-field or the $k$-space of the focused beam via $\boldsymbol{\rho}=-\frac{f}{k_0}(k_x,k_y)$~\cite{novotny_principles_2012}. In the following, we refer to the BFP coordinates as \textit{angularly resolved}. The beam in Eq.~\ref{eq:inc}, schematically shown in Fig.~\ref{fig:focusing}(b), is cylindrically symmetric with respect to the optical axis $z$ and it is linearly polarized in each point of the BFPs. Consequently, it has zero helicity density $K_{\text{in}}^\psi = 0$ everywhere in the beam cross-section. The fields in the proximity of the optical axis in the focal plane ($\rho \approx 0,\,z=0$) produced by the first MO~\cite{novotny_principles_2012,weak2} are given by:
\begin{align} \begin{split}
&\mathbf{E}_{\text{foc}}^\psi \left(\mathbf{r}\right) =A \left( \cos(\psi) \left\{ \rho \hat{\boldsymbol{\rho}} +  \frac{2\imath}{k_{\text{eff}}} \hat{\mathbf{z}} \right\} + \sin(\psi) \frac{k_0 } {k_{\text{eff}}} \rho \hat{\boldsymbol{\varphi}} \right) \\
&\mathbf{H}_{\text{foc}}^\psi \left(\mathbf{r}\right) =B \left( \cos(\psi) \frac{k_0 } {k_{\text{eff}}}   \rho \hat{\boldsymbol{\varphi}} - \sin(\psi)  \left\{ \rho \hat{\boldsymbol{\rho}} + \frac{2\imath}{k_{\text{eff}}} \hat{\mathbf{z}} \right\} \right) 
\text{,}\\
\end{split} \label{eq:foc} \end{align}
where $k_{\text{eff}}$ is the effective wavenumber and $A,\, B\in \mathbb{R},\,B=A \eta^{-1}$ are the proportionality constants~\cite{k_note}. For symmetry reasons, the fields along the optical axis ($\rho=0$) have strictly zero helicity density $K_{\text{foc}}^\psi = 0$ also for the case of $n_1 \neq n_2$ that includes reflection~\cite{novotny_principles_2012}. We assume that the fields in Eq.~\ref{eq:foc} excite an electric-dipole-like spherical gold nanoparticle positioned in the focus $\mathbf{r}_0= \mathbf{r}(\rho,z=0)$. We also assume that this scatterer responds only to the local electric field and the induced electric dipole moment is $\mathbf{p}=\alpha_e \varepsilon_0  \mathbf{E}_{\text{foc}}^\psi(\mathbf{r}_0)\equiv (0,0,p_z)$, where $\alpha_e$ is the electric-dipole polarizability, $k_0=2 \pi/ \lambda$ is the freespace wavenumber, $\lambda$ is the freespace wavelengths and $\varepsilon_0$ is the vacuum permittivity. Importantly, in freespace, the electric-dipole polarizability $\alpha_e$ of a spherical nanoparticle is in quadrature ($\pi/2$ phase-delayed) with its first Mie coefficient $a_1$ and is given by $\alpha_e=(6 \pi \imath  / k_0^3 )a_1$~\cite{bohren1983}. For a dipole moment oriented along the optical axis, the scattered light $\mathbf{E}_{\text{sc}}^\psi$ collected by the second MO is purely radially polarized~\cite{novotny_principles_2012,wozniak_selective_2015,weak1}:
\begin{align} \begin{split}
\mathbf{E}_{\text{sc}}^\psi&=-C D \frac{ \rho }{f} p_z \hat{\boldsymbol{\rho}}=G D a_1 \cos(\psi) \frac{ \rho }{f}  \hat{\boldsymbol{\rho}}\\
\mathbf{H}_{\text{sc}}^\psi&=-C D \frac{ \rho }{f \eta} p_z \hat{\boldsymbol{\varphi}}=G D a_1 \cos(\psi) \frac{ \rho }{f \eta}  \hat{\boldsymbol{\varphi}}
\text{,}\\
\end{split} \label{eq:sc} \end{align}
where $C=\frac{ k_0^2 }{4 \pi \varepsilon_0 f}$, $D= \left[1 - (\rho/f)^2  \right]^{-1/4}$ and $G=\frac{ 3 A }{k_0 f k_\text{eff}}$. Here, also the scattered field has strictly zero helicity density $K_{\text{sc}}^\psi \propto \Im \left( {\mathbf{E}_{\text{sc}}^\psi}^{\ast} \cdot \mathbf{H}_{\text{sc}}^\psi\right) =0$. Surprisingly, the total field $\mathbf{E}_{\text{tot}}^\psi =\mathbf{E}_{\text{inc}}^\psi+ \mathbf{E}_{\text{sc}}^\psi$ in the BFP of the collecting MO acquires a non-zero helicity density $K_{\text{tot}}^\psi$, which can be expressed as the angularly resolved third Stokes parameter $S_3(\rho)$:
\begin{align} \begin{split}
S_3(\rho)&=2\Im \left\{(\mathbf{E}_{\text{tot}}^\psi \cdot  \hat{\boldsymbol{\rho}})^\ast  (\mathbf{E}_{\text{tot}}^\psi \cdot  \hat{\boldsymbol{\varphi}}) \right\}\\
&=2\Im \left\{(\mathbf{E}_{\text{sc}}^\psi \cdot  \hat{\boldsymbol{\rho}})^\ast (\mathbf{E}_{\text{inc}}^\psi  \cdot \hat{\boldsymbol{\varphi}}) \right\}\\
&=- \sin(2\psi) \Im \left\{  a_1\right\} G D \frac{ \rho }{f}  E_\text{in}
\text{.}
\end{split} \label{eq:s3} \end{align} 
Eq.~\ref{eq:s3} highlights the main theoretical result of this manuscript. First, the far-field helicity is generated owing to the superposition between the azimuthally polarized component of the incident beam and the phase-delayed radially polarized component of the scattered light. Secondly, the generation or extinction of helicity is an off-resonance effect governed by $\Im(a_1)$ or $\Re(\alpha_e)$, contrary to the resonant effect of extinction of energy, which is proportional to $\Re(a_1)$ or $\Im(\alpha_e)$~\cite{Nieto2017_05}. Particularly, it is the non-zero phase of $a_1$ that delays the scattered light with respect to the incident beam and results in $S_3(\rho) \neq 0$, as appears in Eq.~\ref{eq:sc} and~\ref{eq:s3}. Thirdly, helicity is extinct by a scatterer positioned at a point where the helicity density of the excitation field $K_{\text{foc}}^\psi$ is zero, in sharp contrast to the mechanism of energy extinction that requires non-zero energy density~\cite{Nieto2017_05,gutsche_optical_2018}. Lastly, it is remarkable that the scattered helicity $\propto \Im( \mathbf{p} \cdot \mathbf{m}^\ast$)~\cite{Nieto2015, Zambrana2016} is also zero, since there is no excited magnetic dipole moment $\mathbf{m} = 0$ for an ideal electric-dipole scatterer. In this case, the extinction of helicity from the incoming beam $\mathcal{W}_\mathcal{H}$ can be calculated as a projection of the induced electric dipole moment on the magnetic focal field $\mathcal{W}_\mathcal{H} \propto -\Re \left\{ \mathbf{p} \cdot \mathbf{H}_{\text{foc}}^{\psi \ast} \left(\mathbf{r}_0\right) \right\}$~\cite{Nieto2015}. On the one hand, substituting the expression for the dipole moment $\mathbf{p}$, we get $\mathcal{W}_\mathcal{H} \propto \Im \left\{ a_1 \mathbf{E}_{\text{foc}}^{\psi } \left(\mathbf{r}_0\right) \cdot  \mathbf{H}_{\text{foc}}^{\psi \ast} \left(\mathbf{r}_0\right) \right\}$, which resembles the definition of the helicity density of the focal fields $K_{\text{foc}}^\psi$ modified by the Mie coefficient $a_1$. On the other hand, substituting the fields in Eq.~\ref{eq:foc}, we get $\mathcal{W}_\mathcal{H} \propto -\sin(2\psi) \Im (a_1) $, showing near-field to far-field (Eq.~\ref{eq:s3}) correspondence. \\
\section{Experiment} 
\begin{figure}[htb!]
  \includegraphics[width=0.5\textwidth]{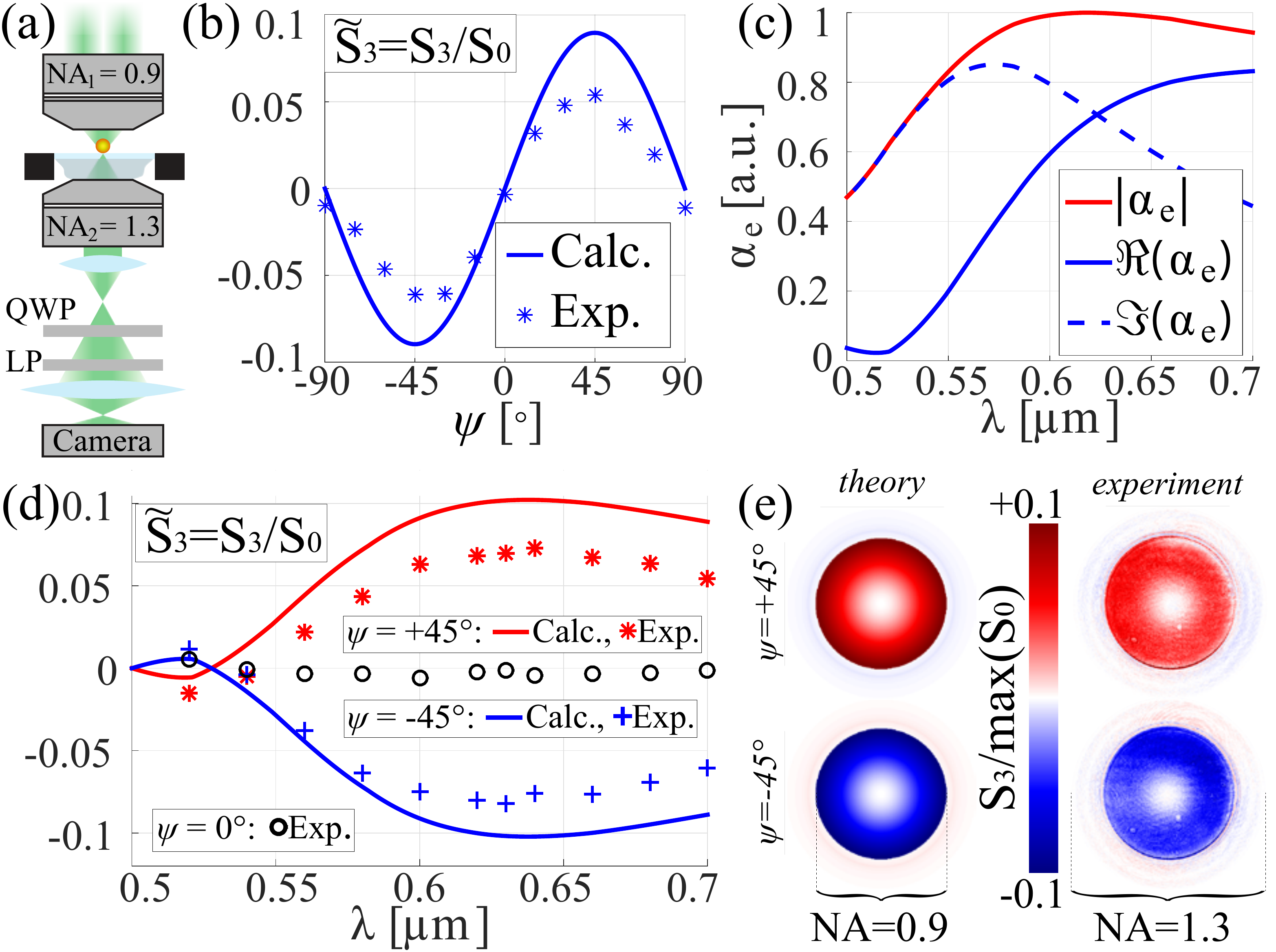}
  \caption{(a) Sketch of the experimental setup. The incident beam is focused onto a gold nanoparticle by a microscope objective (MO). The light propagating in forward direction is collected and collimated by an immersion-type MO and transmitted through an achromatic quarter-wave plate (QWP) and a rotatable linear polarizer (LP). A subsequent lens images the back focal plane of the second MO onto a camera. (b) The calculated (solid line) and measured (markers) averaged normalized value of the third Stokes parameter $\widetilde{S}_3$ as a function of the spiral polarization angle $\psi$ at the wavelength of $\lambda=600\,\mathrm{nm}$. (c) Longitudinal (along the substrate normal) electric-dipole polarizability $\alpha_e$, normalized to its own maximal value, of a spherical gold particle of diameter $d=140 \,\mathrm{nm}$ positioned on a glass substrate with the refractive index $n=1.52$. (d) The calculated (solid lines) and measured (markers) $\widetilde{S}_3$ as a function of wavelength for the spiral polarization angles $\psi=+45\degree$ (red) and $\psi=-45\degree$ (blue). The black circles show the measured $\widetilde{S}_3$ for radial excitation $\psi=0\degree$. (e) Calculated (left column) and measured (right column) angularly resolved $S_3$ parameters for $\psi=+45\degree$ (top row) and $\psi=-45\degree$ (bottom row) at $\lambda = 630\,\mathrm{nm}$, normalized to the maximal value of $S_0$. These images show that $S_3$ is only significant in the angular region containing both the incident and the scattered light (NA$\leq 0.9$), while the scattered light itself ($0.9<$NA$\leq 1.3$) does not contain significant circular polarization.} 
  \label{fig:fig2}
\end{figure}
Our experimental setup, described in detail in our previous works \cite{Banzer2010, Eismann2018}, is shown as a simplified sketch in Fig.~\ref{fig:fig2}(a). We convert an incoming linearly polarized Gaussian beam into a SPCVB using a q-plate~\cite{Marrucci2006} of charge $1/2$. We spatially filter the SPCVB~\cite{karimi_hypergeometric-gaussian_2007} and focus it by the first MO with NA$_1$=0.9. A gold~\cite{Johnson} nanosphere of diameter $d=140\,\mathrm{nm}$ is positioned on the optical axis $z$ in air ($n_1=1$) above a glass substrate ($n_2=1.52$) at $z=-d/2=-70\,\mathrm{nm}$. Mie theory~\cite{bohren1983} predicts that such a nanoparticle in freespace behaves dominantly as an electric-dipole scatterer in the wavelength range $ 520 \,\mathrm{nm} \leq \lambda \leq 700 \,\mathrm{nm}$. The glass substrate is mounted onto a 3D piezo actuator, allowing for precise positioning of the nanoparticle in the focal volume. The transmitted and scattered light are collected by the second confocally aligned index-matched immersion-type MO ($\mathrm{NA}_2=1.3$), while the focal plane ($z=0$) of both MOs constituting a boundary between two media. In our experimental scheme of the nanoparticle on a glass substrate with $\mathrm{NA}_2>\mathrm{NA}_1$, the far-field interference of the incident and scattered light is obtained in the BFP of the second MO in the angular range $|n_2 \sin(\theta_2)| \leq \max\left\{  \sin( \theta_1 ) \right\} =\mathrm{NA}_1$, as schematically depicted in Fig.~\ref{fig:focusing}(a), while for $|n_2 \sin(\theta_2)| > \max\left\{  \sin( \theta_1 ) \right\} =\mathrm{NA}_1$ we collect only the scattered light, which allows us to experimentally verify independently our theoretical predictions in Eq.~\ref{eq:sc} and~\ref{eq:s3}. We image the BFP of the second MO onto an achromatic quarter-wave plate and a linear polarizer to project the field distribution in the BFP onto RCP or LCP. The second lens in Fig.~\ref{fig:fig2}~(a) images the projected BFP intensity distribution $I_\mathrm{RCP}$ or $I_\mathrm{LCP}$ onto a camera, which allows us to measure the far-field angularly resolved Stokes parameters $S_0(\mathbf{k})=I_\mathrm{LCP}+I_\mathrm{RCP}$ and $S_3(\mathbf{k})=I_\mathrm{LCP}-I_\mathrm{RCP}$. We background-correct each measurement by transmitting the excitation beam through the substrate only. To obtain the overall values of $S_0$ and $S_3$ we integrate $S_0(\mathbf{k})$ and $S_3(\mathbf{k})$ across the BFP.\\
To theoretically describe the practical experimental conditions, we include in our theory the actual incident beam and apertures' sizes, the position of the scatterer on the optical axis, the contribution of reflection to the focal fields, dressed electric-dipole and magnetic-dipole polarizabilities of the scatterer~\cite{substrate_halas,substrate_bianisotropy}, Fresnel coefficients and energy conservation factors~\cite{novotny_principles_2012}.\\
First, we perform a measurement at $\lambda = 600\, \mathrm{nm}$ to verify the dependence of the generated far-field helicity on the incident spiral polarization angle $\psi$. In Fig.~\ref{fig:fig2}(b) we plot the normalized average value of the third Stokes parameter $\widetilde{S}_3 \equiv  S_3/S_0$ along with its theoretical value for $-90\degree \leq \psi \leq +90\degree $. For incident radially (TM) ($\psi=0\degree$) and azimuthally (TE) polarized beams ($\psi=\pm 90\degree$) only electric $p_z$ and (a very small, but not strictly zero) magnetic $m_z$ dipoles are symmetry-allowed to be excited~\cite{wozniak_selective_2015}, featuring TM and TE polarized scattered light, respectively, resulting in $\widetilde{S}_3=0$. For $\psi=+45\degree $ and $\psi=-45\degree$ the far-field light is left- and right-handed elliptically polarized, respectively. The experimental results correctly resolve the dependence $\widetilde{S}_3 \propto \sin(2\psi)$, as predicted by Eq.~\ref{eq:s3}.\\
Next, we study the spectrum of $\widetilde{S}_3=S_3/S_0$ at the angles $\psi=\pm 45\degree$. On the one hand, from Eq.~\ref{eq:s3}, we expect the spectrum of $S_3$ to follow $\Im(a_1)$~\cite{Nieto2017_05,gutsche_optical_2018} or, in our experimental configuration, the real part of the longitudinal (along the $z$-axis) electric-dipole polarizability $\Re(\alpha_e)$~\cite{substrate_halas,substrate_bianisotropy}, which is plotted in Fig.~\ref{fig:fig2}(c). On the other hand, $S_0$ is inversely proportional to the extinction of light by the nanoparticle $\propto\,\Im(\alpha_e)$. As a result, both $\Re(\alpha_e)$ and $\Im(\alpha_e)$ influence the spectrum of $\widetilde{S}_3=S_3/S_0$. In Fig.~\ref{fig:fig2}(d) we plot the resulting calculated and the experimentally obtained values of $\widetilde{S}_3$ for $\psi=+ 45\degree$ and $\psi=- 45\degree$ in red and blue color, respectively. Fig.~\ref{fig:fig2}(d) confirms the generation of non-zero far-field helicity with the sign of $\widetilde{S}_3$ being dependent on $\psi$, in agreement with our theoretical model. We were also able to experimentally resolve the sign flip of $\widetilde{S}_3$ at around $\lambda \approx 530\, \mathrm{nm}$. This sign flip does not appear in $\Re(\alpha_e)$ in Fig.~\ref{fig:fig2}(c) and it is a result of the wavelength-dependent phase of the reflected incident field exciting the scatterer. Additionally, we perform a calibration experiment with radially polarized excitation ($\psi=0\degree$), where the expected value of $\widetilde{S}_3$ is zero, shown as black circles in Fig.~\ref{fig:fig2}(c).\\
Lastly, we experimentally confirm that in the angular region $|n_2 \sin(\theta_2)| > \max\left\{  \sin( \theta_1 ) \right\} =\mathrm{NA}_1$, where the scattered light does not interfere with the incident beam, the third Stokes parameter $S_3$ of the far-field light is close to zero. In Fig.~\ref{fig:fig2}(e), we plot the theoretically calculated (left column) and experimentally recorded (right column) angularly resolved $S_3$ parameters for $\psi = +45 \degree$ (top row) and $\psi = -45 \degree$ (bottom row) at $\lambda = 630\,\mathrm{nm}$, normalized to the maximum value of $S_0$. Fig.~\ref{fig:fig2}(e) confirms that only within the angular range corresponding to the interference of the incident and scattered light (NA$ \leq 0.9$) we observe a significant value of $S_3$. On the contrary, in the angular range $ 0.9 \leq $NA$ \leq 1.3$, corresponding to the scattered light only, we obtain negligible values of $S_3 \approx 0$. The small (but non-zero) residual values in the angular range $ 0.9 \leq $NA$ \leq 1.3$ originate from the contribution of a small (but not strictly zero) magnetic dipole moment~\cite{Eismann2018} supported by the gold nanoparticle.
\section{Discussion and Conclusion}
The observed effect can be understood by considering the helicity conservation laws, symmetry and the duality properties of our system~\cite{Ivan2012,Cameron_helicity2012,Ivan2013,Bliokh2013,Nieto2015}. Electric-dipole scatterers break the electromagnetic duality symmetry by reacting selectively to electric field only. Systems that break the electromagnetic duality symmetry do not conserve helicity, i.e., they may change the average far-field degree of circular polarization~\cite{Ivan2013}, which we have experimentally confirmed. These effects effectively mimic chiral light-matter interaction. Additionally, our system is also rotationally invariant, meaning that the total angular momentum of light $J_z$ must be conserved. As a result, the emerging spin angular momentum in the region of NA$ \leq 0.9$ must be compensated by the generation of orbital angular momentum~\cite{Bliokh2014}, i.e., the circularly polarized components of the beam shown in Fig.~\ref{fig:fig2}(e) have a helical phase distribution as a direct consequence of the phase-shifted superposition of azimuthal polarization of the excitation beam and radial polarization emitted by the nanoparticle~\cite{Eismann2018}. Finally, the presented effect is determined by the longitudinal field components, accentuating their importance in the description of light-matter interaction~\cite{long_novotny,blokh_soi_nanoprobing_2010,BanzerWheel2013,aiello_transverse_2015,Nieto2017_05,long_chirality,fano_yuri_2018,wozniak_interaction_2019,nechayev_orbital--spin_2019}.\\
In conclusion, we have theoretically and experimentally shown that scattering of a focused cylindrical vector beam with spiral polarization by an electric-dipole scatterer positioned on the optical axis generates far-field circular polarization. Notwithstanding that the incident beam, the focal fields, as well as the scattered light have zero optical chirality. The effect originates from the electromagnetic duality symmetry breaking by the scatterer, which selectively responds to the electric field only. It can be conveniently explained in terms of the superposition of the transverse electric (azimuthal) component of the incident beam and the phase-delayed transverse magnetic (radial) component of the scattered light. Utilizing a substrate supporting the scatterer allowed us to separate the angular region where the transmitted far-field interferes with the scattered light from the angular region that contains the scattered light only, facilitating experimental observation of our theoretical predictions. Our experiment and the simple theoretical model shed light on chiral light-matter interactions, helicity conservation laws in electromagnetism, emphasize the role of duality symmetry in optics.
\begin{acknowledgments}
We gratefully acknowledge fruitful discussions with Pawe{\l} Wo{\'z}niak, Martin Neugebauer, J{\"o}rg S.~Eismann and Gerd Leuchs. The authors thank Eduard Butzen for providing the scanning electron microscopy images of the sample.
\end{acknowledgments}
\bibliography{bib}
\end{document}